\documentclass[12pt,a4]{article}
\begin{document}
\renewcommand{\baselinestretch}{1.0}
\newcommand{\Section}{\setcounter{equation}{0}\section}
\begin{center}
STRUCTURE OF EXTENDED LOOP WAVE FUNCTION IN QUANTUM GRAVITY AND OPERATOR FORMALISM

$\vspace{2cm}$

\footnotesize{$\rm{M. SAIT{O}^{\sharp,\P}, H. NOD{A}^{+,\|}\ and\ T. TASHIR{O}^{\S,}}$

$\P Graduate\ School\ of\ Science\ and\ Engineering,\ Ibaraki\ University,\ Mito\ 310-8512,\ Japan$
$\| Department\ of\ Mathematical\ Science,\ Ibaraki\ University,\ Mito\ 310-8512,\ Japan$
$\dag Department\ of\ Computer\ Simulation,\ Okayama\ University\ of\ Sciense,\ Okayama\ $ $700-0005,\ Japan$

$\sharp:\ saito@serra.sci.ibaraki.ac.jp$

$+:\ noda@mx.ibaraki.ac.jp$

$\S:\ tashiro@sp.ous.ac.jp$
}

\end{center}

$\vspace{0.5cm}$

\small{The structure of extended loop wave function is investigated in terms of the operator formalism. It is found that the extended loop wave function is characterized by the family number and classified by the partition of the family number. It is pointed out that the constraints to the extended loop function in quantum gravity exhibit a hierarchy structure.}

$\vspace{0.5cm}$

\small{$Keywords$:$\ $Quantum gravity;$\ $Extended loop wave function;$\ $Operator formalism.}

$\vspace{0.5cm}$
\small{\section{Introduction}}

\normalsize{
A new set of canonical variables for the Hamiltonian treatment of general relativity was introduced 
by Ashtekar \cite{Ashtk86}. The new set of canonical variables are the triads $E_i^{ax} $ 
(the projections of the tetrads onto a three-surface) and a complex $SU(2)$ connection $A_{ax}^i$. 
The introduction has opened new possibilities of achieving a Dirac canonical quantization of the gravitational field. 
It was shown that there exists a solution to the Hamiltonian constraint with a cosmological constant $\Lambda $, 
which is also diffeomorphism invariant, given by the exponential of the Chern-Simons form \cite{Ashtk90}. 
The loop representation of this solution can be written as a knot polynomial in the cosmological constant,
\begin{eqnarray}
{\psi_{AK}}=\sum_{m=0}^{\infty}(\frac{\Lambda }{6})^mK_m= e^{\frac{\Lambda }{6}a_1}\tilde{J_{\Lambda }}\nonumber
\end{eqnarray}
where $K_m$ denotes the $m$-th coefficient of the Kauffman bracket and $\tilde{J_{\Lambda }}$ the Jones polynomial. 
The function $a_1$  is proportional to Gauss self-linking number \cite{E. Guadagnini}.

  Recently, the systematic method to obtain analytic expressions of the diffeomorphism constraints, has been 
developed \cite{J. Griego,D. Shao}. This approach to get extended knot invariants has particular significance for 
the search of the gravitational quantum states, whose candidates must satisfy the diffeomorphism constraint, 
Hamiltonian constraint and the Mandelstam identities \cite{S. Mandelstam}.
  
In this article, the gravitational quantum states are investigated in the framework of the extended loop 
representation \cite{C. Di Bartolo94,C. Di Bartolo95,J. GriegoB467}. An operator formalism for the extended 
loop wave function is formulated. It is possible to obtain systematically analytic expressions of the extended 
loop wave functions. It is discussed that they satisfy the Mandelstam identities, the diffeomorphism constraint 
and Hamiltonian constraint.

The article is organized as follows: in Section 2, the extended loop wave functions are reformulated by the bra-ket formalism. In Section 3, the operator formalism for propagator is investigated. The concept of the family number is introduced to characterize the structure of the wave functions. It is found that the extended loop wave function is characterized by the family number and classified by the partition of the family number. In Section 4, the extended loop wave functions in quantum gravity are analyzed. The Mandelstam identities, the diffeomorphism and Hamiltonian constraints are discussed. It is pointed out that the constraints to the extended loop function in quantum gravity exhibit a hierarchy structure. In Section 5, conclusion and discussion are given.
}

\small{\section{Extended loop representation in quantum gravity}}
\normalsize{
The quantum state vector of the universe is written as $|\psi>$. The ket $|\mu > $ is introduced as the following infinite basic vectors;
\begin{eqnarray}|\mu>=\{|\mu_1\cdots \mu_r>;r=0,1,2,\cdots ,\infty \}
\end{eqnarray}
where $\mu_i$ represents the pair of variables $(a_i , x_i )$ with vector index $a_i$  and space point $x_i$. The number of indexes of the set $\mu$  is $n(\mu )$. The ket $|\mu>$ is the eigenstate of the number operator of indexes $\hat{N}$:$\hat{N}|\mu>=n(\mu)|\mu>$. The ket $|\mu>$ forms a complete set
\begin{eqnarray}
\sum_{\mu}|\mu><\mu|=\sum_{r=0}^{\infty }|\mu_1\cdots \mu_r><\mu_1\cdots \mu_r|=1.\nonumber 
\end{eqnarray}
Also, the overline infinite basic vector $|\bar{\mu}>$   is defined as 
\begin{eqnarray}
|\bar{\mu}>=\{|\overline{\mu_1\cdots \mu_r}>\equiv (-1)^r|\mu_1\cdots \mu_r>;r=0,1,2,\cdots ,\infty \}
\end{eqnarray}
where the factor $(-1)^r$  denotes the parity of the ket $|\mu_1\cdots \mu_r>$.
\footnote[1]{The inverse operator $\hat{I}$ and the parity operator $\hat{\pi}$ are introduced as follows:
\begin{eqnarray}
\hat{\pi}|\mu_1\cdots \mu_r>=|\mu_r\cdots \mu_1>,\ \ \hat{\pi}|\mu_1\cdots \mu_r>=(-1)^r|\mu_1\cdots \mu_r>.\nonumber
\end{eqnarray}
The ket $|\bar{\mu}>$ is related to the ket $|\mu>$ as $|\bar{\mu}>=\hat{T}|\mu>$ where $\hat{T}=\hat{I} \hat{\pi}$. The eigenvalue of $\hat{T}$ are $\pm{ 1}$ since ${\hat{T}}^2 =1$.}

The extended loop wave function are constructed by the multi-vector density fields and the propagators. The multi-vector density fields are elements of the extended loop group and have the following form: 
\begin{eqnarray}
{\bf X}=\{ X^{\mu_1\cdots \mu_r};r=0,1,2,\cdots \infty \}
\end{eqnarray}
with
\begin{eqnarray}
X^{\mu_1\cdots \mu_r}=<{\bf X}|\mu_1\cdots \mu_r>.\nonumber
\end{eqnarray}
The number of paired indexes defines the rank of the multi-vector field. 

The extended loop wave function $\psi ({\bf X})$ is linear in the multi-vector fields and are written in general as follows:
\begin{eqnarray}
\psi ({\bf X})=\sum_{r=0}^{\infty }X^{\mu_1\cdots \mu_r}\psi_{\mu_1\cdots \mu_r}=<{\bf X}|\psi>
\end{eqnarray}
with
\begin{eqnarray}
\psi_{\mu_1\cdots \mu_r}=<\mu_1\cdots \mu_r|\psi>\nonumber
\end{eqnarray}
where a generalized Einstein convention is assumed and $\psi_{\mu_1\cdots \mu_r}$ denotes the propagator. 
This linearity of the extended loop wave function is guaranteed by the property of the extended holonomy 
in $X^{\mu_1\cdots \mu_r}$ \cite{C. Di Bartolo95}. 

In quantum gravity, the propagator $\psi_{\mu_1\cdots \mu_r}$ should satisfy a set of symmetry properties, that is, 
the Mandelstam identities \cite{S. Mandelstam}. They are
\begin{eqnarray}
({\rm I})\ \ \ <\mu_1\cdots \mu_r|\psi>=<(\mu_1\cdots \mu_r)_c|\psi>,
\end{eqnarray}
\begin{eqnarray}
({\rm II})\ \ \ <\mu_1\cdots \mu_r|\psi>=<\overline{\mu_1\cdots \mu_r}|\psi>,
\end{eqnarray}
\begin{eqnarray}
({\rm III}) <\mu_1\cdots \mu_k \mu_{k+1}\cdots \mu_r|\psi>+<\overline{\mu_1\cdots \mu_k}\mu_{k+1}\cdots \mu_r|\psi>\nonumber\\
=\frac{1}{k}\{<(\mu_1\cdots \mu_k)_c \mu_{k+1}\cdots \mu_r|\psi>+<\overline{(\mu_1\cdots \mu_k)_c}\mu_{k+1}\cdots \mu_r|\psi>\}  
\end{eqnarray}
for all $k$ where $c$ indicates the cyclic combination of indexes. 

The overline multi-vector density fields $\bar{{\bf X}}$ are defined as 
\begin{eqnarray}
\bar{{\bf {X}}}=\{{\bar{X}}^{\mu_1\cdots \mu_r}\equiv {(-1)^r}X^{\mu_r\cdots \mu_1};r=0,1,2,\cdots \infty \}\nonumber
\end{eqnarray}
with
\begin{eqnarray}
{\bar{X}}^{\mu_1\cdots \mu_r}=<{\bar{\bf X}}|\mu_1\cdots \mu_r>.\nonumber
\end{eqnarray}
It is noted that $|\bar{{\bf X}}>=\hat{T}|{\bf X}>$. Among $|\mu>$, $|\bar{\mu}>$, $|{\bf X}>$ and $|\bar{{\bf X}}>$, the following relations hold:
\begin{eqnarray}
<{\bf X}|\mu>=<\bar{{\bf X}}|\bar{\mu}>,\ <\bar{{\bf X}}|\mu>=<{\bf X}|\bar{\mu}>.
\end{eqnarray}
If Eq.(6) holds, we obtain $\psi({\bf X})=\psi(\bar{{\bf X}})=\psi({\bf R})$  because of Eq.(8) where ${\bf R}$ $=$ $({\bf X}$ $+\bar{{\bf X}})/2$.

Furthermore, the extended loop wave function must satisfy the following constraints:

1) Diffeomorphism constraint
\begin{eqnarray}
<{\bf R}|\hat{C}_{ax}|\psi>=\sum_{r=0}^{\infty }<{\bf R}|\hat{C}_{ax}|\mu_1\cdots \mu_r><\mu_1\cdots \mu_r|\psi>=0,
\end{eqnarray}
\begin{eqnarray}
<{\bf R}|\hat{C}_{ax}|\mu_1\cdots \mu_r>={\cal{F}}_{ab}^{\mu_1}(x){R}^{(bx\mu_2 \cdots \mu_r)_c}+{\cal{F}}_{ab}^{{\mu}_{1}{\mu}_{2}}(x){R}^{(bx\mu_3\cdots \mu_r)_c}\nonumber
\end{eqnarray}
with
\begin{eqnarray}
{\cal{F}}_{ab}^{{\mu}_{1}}(x)={\delta}_{a}^{{a}_{1}}{\delta}_{b}^{{d}}{\partial}_{d}\delta({x}_{1}-x),\nonumber
\end{eqnarray}
\begin{eqnarray}
{\cal{F}}_{ab}^{{\mu}_{1}{\mu}_{2}}(x)={\delta}_{a}^{{a}_{1}}{\delta}_{b}^{{a}_{2}}\delta({x}_{1}-x)\delta({x}_{2}-x)\nonumber
\end{eqnarray}
where ${\hat{C}}_{ax}$ denotes the diffeomorphism constraint operator. This wave function satisfies Eq.(9) and thus is knot invariant.

2) Hamiltonian constraint

 a) Vacuum Hamiltonian $({\hat{\cal{H}}}_{0}(x))$; 
\begin{eqnarray}
<{\bf R}|{\hat{\cal{H}}}_{0}(x)|\psi>=\sum_{r=0}^{\infty }<{\bf R}|{\hat{\cal{H}}}_{0}(x)|\mu_1\cdots \mu_r><\mu_1\cdots \mu_r|\psi>=0,
\end{eqnarray}
\begin{eqnarray}
<{\bf R}|{\hat{\cal{H}}}_{0}(x)|\mu_1\cdots \mu_r>=2\{{\cal{F}}_{ab}^{{\mu}_{1}}(x){R}^{(ax,bx){\mu}_{2}\cdots{\mu}_{r}}+{\cal{F}}_{ab}^{{\mu}_{1}{\mu}_{2}}(x){R}^{(ax,bx){\mu}_{3}\cdots{\mu}_{r}} \} \nonumber
\end{eqnarray}
with
\begin{eqnarray}
{R}^{(ax,bx){\mu}_{1}\cdots{\mu}_{r}} \equiv \sum_{k=0}^{r}{R}^{(ax{\mu}_{1}\cdots{\mu}_{k}bx\overline{{\mu}_{k+1}\cdots{\mu}_{r}})_{c}}. \nonumber
\end{eqnarray}
The sequences ${\mu}_{1}\cdots{\mu}_{0}$ and ${\mu}_{r}\cdots{\mu}_{r+1}$ for $k=0$ and $k=r$ are assumed to be the null set of indexes.

 b) Hamiltonian with $\Lambda$ $({\hat{\cal{H}}}_{\Lambda}(x))$ ;

\begin{eqnarray}
<{\bf R}|{\hat{\cal{H}}}_{\Lambda}(x)|\psi>=0
\end{eqnarray}
with
\begin{eqnarray}
{\hat{\cal{H}}}_{\Lambda}(x)={\hat{\cal{H}}}_{0}(x)+\frac{\Lambda}{6}{\rm detq} \nonumber
\end{eqnarray}
where q is the three metric. The cosmological term is given as
\begin{eqnarray}
<{\bf R}|{\rm detq}|\psi>&=&2{\epsilon}_{abc}\sum_{r=0}^{\infty} \sum_{\rho ,\nu ,\mu} <{\rho\nu\mu}|\psi>{(-1)}^{n(\mu)}\{{R}^{(ax{\mu}^{-1}bx\nu cx\rho)_{c}} \nonumber\\
&+&{R}^{(ax\rho bx{\mu}^{-1}cx\nu)_{c}}+{R}^{(ax\rho bx\nu cx{\mu}^{-1})_{c}}\} \nonumber
\end{eqnarray}
where $n(\rho)+n(\nu)+n(\mu)=r$, $\rho={\mu}_{1}\cdots{\mu}_{i}$, $\nu={\mu}_{i+1}\cdots{\mu}_{j}$, 
$\mu={\mu}_{j+1}\cdots{\mu}_{r}$ and $0\leq i\leq j\leq r$ \cite{D. Shao03}. Also, ${\mu}^{-1}$ is ${\mu}_{r}\cdots{\mu}_{1}$ if $\mu={\mu}_{1}\cdots{\mu}_{r}$.

}
\vspace*{0.5cm}

\small{\section{Extended loop wave function and operator formalism}}
\small{\subsection{ Propagator operators}}
\normalsize{
First, we assume the linearity of the extended loop wave function of Eq.(4).

Secondly, we assume that $ |\psi>=\tilde{G}|0>$ where $\tilde{G}$ denotes the propagator operator and $|0>$ is the vacuum state. The propagator $\psi_{\mu_1\cdots \mu_r}$ is defined as follows:
\begin{eqnarray}
\psi_{\mu_1\cdots \mu_r}=<\mu_1\cdots \mu_r|\tilde{G}|0>.
\end{eqnarray}
From Eq.(4), the extended loop wave function is rewritten as 
\begin{eqnarray}
\psi({\bf X})=\sum_{r=0}^{\infty }X^{\mu_1\cdots \mu_r}<\mu_1\cdots \mu_r|\tilde{G}|0>=<{\bf X}|\tilde{G}|0>.
\end{eqnarray}
It is noted that the ordering of $\mu_1\cdots \mu_r$ in $X^{\mu_1\cdots \mu_r}$ is fixed. 
 
Thirdly, it is assumed that the operator $\tilde{G}$ has the following properties:
\begin{enumerate}
\item[1)]	The operator $\tilde{G}$ is built up by the basic operator with the rank $r$; $G_r $ $(r=2,3,\cdots ,\infty)$ 
\item[2)]	The basic operator $G_r$ has the coupling constant $\lambda_r ={\lambda}^{r-1} $. That is $\tilde{G} =\lambda_rG_r $.
\item[3)]	The basic operators $G_s$ and $G_t$ are commutable; $G_sG_t=G_tG_s$.
\item[4)]  The rank of the product of $G_s$ and $G_t$ is $(s+t)$.
\item[5)]  The matrix element of $G_s$  is given as:
\begin{eqnarray}
<{\mu_1\cdots \mu_r}|G_s|0>=\delta_r^s<\mu_1\cdots \mu_r|G_s|0>.
\end{eqnarray}
\end{enumerate}
From now on, the ordering of $G_s$ and $G_t$ is fixed as $G_sG_t $ if $s\le t$ from 3). The coupling constant $\lambda$ is an arbitrary parameter for the vacuum Hamiltonian, but an adequate one for the Hamiltonian with $\lambda$, that is $\lambda\propto \Lambda$. We observe here that the power of the coupling constant $\lambda$ of $G_r$ differs from its rank $r$.

The operator $\tilde{G}$ is written as the products of basic operators as
\begin{eqnarray}
\prod _{i=1}^m(\lambda_{i+1}G_{i+1})^{f_i}=\lambda^{F}\prod _{i=1}^m(G_{i+1})^{f_i} \nonumber
\end{eqnarray}
where $F=\sum_{i=1}^m if_i$ and $f_i$ is non-negative integer. We notice the power of the coupling constant $\lambda$. If the set $(f_1,\cdots ,f_m)$ is given, we may introduce the family number $m$ as
\begin{eqnarray}
m=\sum_{i=1}^{m}if_i.
\end{eqnarray}
The set $(f_1,\cdots ,f_m)$ is determined by the partition of $ m$;$ [\tau]=[\tau_1$,$\cdots ,\tau_m]$.\footnote[2]{The set of $[\tau_1,\cdots ,\tau_m]$ is give by $\tau_1=f_1+\cdots +f_m$, $\tau_2=f_2+\cdots +f_m$, $\tau_3=f_3+\cdots +f_m$,$\cdots$ and $\tau_m=f_m$. Then, from Eq.(15), $\tau_1+\cdots +\tau_m=1$ and $\tau_1\ge \cdots \ge \tau_m\ge 0$. Thus, the set of $[\tau_1,\cdots ,\tau_m]$ is the partition of $m$. Then we have $F_m =(\tau_1-\tau_2, \tau_2-\tau_3,\cdots ,\tau_m)$.} Then, the operator $\tilde{G}$ is written as 
\begin{eqnarray}
\tilde{G}=\sum_{m=0}^{\infty }{\lambda}^m\sum_{F_m}G_{\bar{r}}(F_m)
\end{eqnarray}
with
\begin{eqnarray}
G_{\bar{r}}(F_m)=\prod _{i=1}^m(G_{i+1})^{f_i}\nonumber
\end{eqnarray}
where $F_m =(f_1,\cdots ,f_m)$ and the sum with respect to $F_m$ is over all the partition of $m$. Also, $\bar{r}$ is defined as
\begin{eqnarray}
\bar{r}=\sum_{i=1}^{m}(i+1)f_i.
\end{eqnarray}
The rank of $G_{\bar{r}}(F_m)$ is $\bar{r}$. 

The wave function of Eq.(13) is decomposed in terms of the power of the coupling constant $\lambda$ as follows;
\begin{eqnarray}
<{\bf X}|\psi>=\sum_{m=0}^{\infty }{\lambda}^m<{\bf X}|\psi_{m}>
\end{eqnarray}
with
\begin{eqnarray}
<{\bf X}|\psi_m>=\sum_{F_m}X^{\mu_1\cdots \mu_{\bar{r}}}<\mu_1\cdots \mu_{\bar{r}}|G_{\bar{r}}(F_m)|0>.
\end{eqnarray}
where $<{\bf X}|\psi_m>$ is the wave function with the family number $m$. Eq.(18) suggests that the wave function with the family number $m$ is constructed by quantum states with the rank $r$($ n < \bar{r} < N :n=m+1$ and $N=2m$). From Eq.(13)and Eq.(16), the extended loop wave function is compactly rewritten as 
\begin{eqnarray}
\psi({\bf X})=<{\bf X}|\prod_{l=2}^{\infty }(1-\lambda_lG_l)^{-1}|0>.
\end{eqnarray}

From Eq.(18), we remark that the extended loop wave functions have the following structure: 
\begin{enumerate}
\item[(1)] The extended loop wave functions are characterized by the family number $m$.
\item[(2)] The extended loop wave function with $m$, is decomposed into the classes in terms of the partition of $m$; $[\tau]=[\tau_1$,$\cdots ,\tau_m]$. 
\item[(3)] The extended loop wave functions are structured according to a sequential arrangement of basic propagators.
\end{enumerate}
}

\vspace*{0.5cm}

\small{\subsection{ Propagator operator and new basic vectors}}
\normalsize{
From Eq.(19), the propagator with the family number $m$ and the rank $\bar{r}$ is given as 
\begin{eqnarray}
\psi_{\mu_1\cdots \mu_{\bar{r}}}^{m}=<\mu_1\cdots \mu_{\bar{r}}|G_{\bar{r}}(F_m)|0>.\nonumber
\end{eqnarray}

The propagator must satisfy the cyclic property from Eq.(5). Also, the operator $G_{\bar{r}}(F_m)$ is rewritten as the products of the basic propagator operators with various ranks. It is necessary to factorize the matrix element $<\mu|G_{\bar{r}}(F_m)|0>$ in order to calculate them. We assume the following points:
\begin{enumerate}
\item[(a)]	The cyclic property of the basic propagator $G_r$.
\item[(b)]	The uniqueness of the matrix element of $G_r$.
\item[(c)]	The factorization of the matrix element of $G_{\bar{r}}(F_m)$.
\end{enumerate}
The ket $|\mu>$ is ordered but not cyclic. We introduce the new bases $|\mu_d>$ instead of $|\mu>$ as follows:
\begin{eqnarray}
|\mu_d>=\{|[\mu_1\cdots \mu_{r}]>;r=0,1,\cdots ,\infty \}
\end{eqnarray}
where $[\mu_1\cdots \mu_{r}]$ denotes the set whose elements are represented by the product of cycles of the variables $\mu_1\cdots \mu_{r}$ according to the requirement that the $q$-term cycle $(\mu_{i_{1}}\cdots \mu_{i_{q}})$ is constrained by the ordering of indexes with $i_1<\cdots <i_q$, that is the "ordered" cycle. This requirement guarantees the assumption (b).\footnote[3]{ For example,
\begin{eqnarray}
[\mu_1\mu_2\mu_3]=\{(\mu_1)(\mu_2)(\mu_3),(\mu_1)(\mu_2\mu_3),(\mu_2)(\mu_1\mu_3),(\mu_3)(\mu_1\mu_2),(\mu_1\mu_2\mu_3)\}.\nonumber
\end{eqnarray}
It is noted that the cycle $(\mu_1\mu_3\mu_2)$ is excluded.} In this case, the states are degenerate.

 Using the ordered cycle of variables $\mu_1\cdots \mu_{\bar{r}}$, we may define the matrix element of $G_{\bar{r}}(F_m)$ as follows:
\begin{enumerate}
\item[(1)] The basic propagator operator with the rank $\bar{r}$, $G_r $:
\begin{eqnarray}
<(\mu_1\cdots \mu_{q})|G_{\bar{r}}|0>=\delta_{q}^{\bar{r}}<(\mu_1\cdots \mu_{\bar{r}})|G_{\bar{r}}|0>\equiv \delta_{q}^{\bar{r}}G_{\mu_1\cdots \mu_{\bar{r}}}
\end{eqnarray}
where $G_{\mu_1\cdots \mu_{\bar{r}}}$ is cyclic symmetric function:$ G_{\mu_1\cdots \mu_{\bar{r}}}= G_{(\mu_1\cdots \mu_{\bar{r}})_c}$. 
\item[(2)] The product of $G_s$ and $G_t$( $s\le t$ and $\bar{r}=s+t$):
\begin{eqnarray}
& &<(\mu_1\cdots \mu_{p})(\mu_{p+1}\cdots \mu_{p+q})|G_sG_t|0>\nonumber\\
& &=\delta_{p}^{s}\delta_{q}^{t}<(\mu_1\cdots \mu_{s})(\mu_{s+1}\cdots \mu_{s+t})|G_{s}G_t|0>\nonumber\\ 
& &\equiv \delta_{p}^{s}\delta_{q}^{t}G_{\mu_1\cdots \mu_{s}}G_{\mu_{s+1}\cdots \mu_{s+t}}.
\end{eqnarray}
\end{enumerate}
Furthermore, we may consider the generarization of Eq.(22) and Eq.(23) in the similar way. 

Using the complete set of  $|\mu_d>$, we obtain
\begin{eqnarray}
<{\bf X}|\psi_m>=\sum_{F_m}X^{\mu_1\cdots \mu_{\bar{r}}}\sum_{\mu_d}\delta_{\bar{r}}^{n(\mu_d)}D_{\mu_1\cdots \mu_{\bar{r}}}^{\mu_d}<\mu_d|G_{\bar{r}}(F_m)|0>
\end{eqnarray}
where $D_{\mu_1\cdots \mu_{\bar{r}}}^{\mu_d}$ is the expansion coefficient and the representative $<\mu|\mu_d>$ is given by $<\mu|\mu_d>=\delta_{n(\mu)}^{n(\mu_d)}D_{\mu}^{\mu_d}$. From Eq.(23), Eq.(24) reduces to
\begin{eqnarray}
<{\bf X}|\psi_m>=\sum_{F_m}X^{\mu_1\cdots \mu_{\bar{r}}}\sum_{[\mu_1\cdots \mu_{\bar{r}}]}D_{\mu_1\cdots \mu_{\bar{r}}}^{[\mu_1\cdots \mu_{\bar{r}}]}<[\mu_1\cdots \mu_{\bar{r}}]|G_{\bar{r}}(F_m)|0>.\nonumber
\end{eqnarray}
From the definition of the matrix element of $G_{\bar{r}}(F_m)$, we obtain
\begin{eqnarray}
<{\bf X}|\psi_m>=\sum_{F_m}X^{\mu_1\cdots \mu_{\bar{r}}}\sum_{\tilde{P}_{\bar{r}}{[\{F_m\}]}}D_{\mu_1\cdots \mu_{\bar{r}}}^{\{F_m\}}<\{F_m\}|G_{\bar{r}}(F_m)|0>
\end{eqnarray}
with
\begin{eqnarray}
\{F_m\}&=&(\mu_1\mu_2)\cdots (\mu_{2f_1-1}\mu_{2f_1})\nonumber\\
&\times& (\mu_{2f_1+1}\mu_{2f_1+2}\mu_{2f_1+3})\cdots (\mu_{2f_1+3f_2-2}\mu_{2f_1+3f_2-1}\mu_{2f_1+3f_2})\nonumber\\
& &\ldots \nonumber\\ 
&\times &(\mu_{2f_1+\cdots +mf_{m-1}+1}\cdots \mu_{2f_1+\cdots +mf_{m-1}+m+1})\nonumber\\
& &\cdots (\mu_{2f_1+\cdots +(m+1)f_{m}-m}\cdots \mu_{2f_1+\cdots +(m+1)f_{m}})
\end{eqnarray}
where $f_i$ denotes the number of the $(i+1)$-term cycle $(\mu_{j_1}\cdots \mu_{j_{f_i}})$ and the cyclic structure of $\{F_m \}$ is characterized by $K=2^{f_1}3^{f_2}\cdots(m+1)^{f_m}$. Then the sum with respect to $\tilde{P}_{\bar{r}}{[\{F_m\}]}$ is over all the permutations by the products of the ordered cycles in the set $\mu_1\cdots \mu_{\bar{r}}$.

The result from Eq.(25) is formally obtained by using Eq.(24) and the following relations:
\begin{enumerate}
\item[(1)] For $G_r$,
\begin{eqnarray}
<\mu_d|G_{r}|0>=\delta_{\mu_d}^{(\mu_1\cdots \mu_{r})}<(\mu_1\cdots \mu_{r})|G_{r}|0>.
\end{eqnarray}
\item[(2)] For $G_sG_t$,
\begin{eqnarray}
<\mu_d|G_{s}G_{t}|0>&=&\sum_{\tilde{P}_{s+t}[(\mu_1\cdots \mu_{s})(\mu_{s+1}\cdots \mu_{s+t})]}\delta_{\mu_d}^{(\mu_1\cdots \mu_{s})(\mu_{s+1}\cdots \mu_{s+t})}\nonumber\\
&\times& <(\mu_1\cdots \mu_{s})(\mu_{s+1}\cdots \mu_{s+t})|G_{s}G_{t}|0>
\end{eqnarray}
where the sum is over all the permutations in terms of the product of the $s$-term cycle $(\mu_1\cdots \mu_s)$ and the $t$-term cycle $(\mu_{s+1}\cdots \mu_{s+t})$         in the set $\mu_1\cdots \mu_{s+t}$. The number of the elements is $n(s,t)=(s+t)!/s!t!$ for $s<t$ and $(2s)!/(s!)^22!$ for $s=t$. 
\item[(3)] For $G_{\bar{r}}(F_m)$,
\begin{eqnarray}
<\mu_d|G_{\bar{r}}(F_m)|0>=\sum_{\tilde{P}_{\bar{r}}[\{F_m\}]}\delta_{\mu_d}^{\{F_m\}}<\{F_m\}|G_{\bar{r}}(F_m)|0>.
\end{eqnarray}
The number of the elements is given as
\begin{eqnarray}
n(F_m)=\frac{[2f_1+3f_2+\cdots +(m+1)f_m]!}{[2!]^{f_1}[3!]^{f_2}\cdots [(m+1)!]^{f_m}f_1!f_2!\cdots f_m!}.
\end{eqnarray}
\end{enumerate}

The extended loop wave functions with $m=1,2,3$ and $4$ are given in Table \ref{tt:1}. These wave functions exhibit a skelton structure. 
\begin{table}[b]
\caption{The extended loop wave functions with $m=1, 2, 3$ and $4$}
\label{tt:1}
\begin{center}
\begin{tabular}{|c| c|c| c| c| c|c|}
\hline
$ $ &\small{Partition of $m$} &$F_m$&\small{Operator} &\small{Rank} &\small{Cyclic}& \\
$<{\bf X}|\psi _m>$& $[\tau_1,\cdots, \tau_m]$&$(f_1,\cdots ,f_m)$&$\tilde{G}$&$\bar{r}$&\small{structure} $K$&$n(F_m)$\\
\hline
$<{\bf X}|\psi _1>$&[1]&(1)&$G_2$&2&$2^1$&1\\
\hline
$<{\bf X}|\psi _2>$&[2,0]&(2,0)&$(G_2)^2$&4&$2^2$&3\\
$ $&[1,1]&(0,1)&$G_3$&3&$3^1$&1\\
\hline
$<{\bf X}|\psi _3>$&[3,0,0]&(3,0,0)&$(G_2)^3$&6&$2^3$&15\\
$ $&[2,1,0]&(1,1,0)&$G_2G_3$&5&$2^13^1$&10\\
$ $&[1,1,1]&(0,01)&$G_4$&4&$4^1$&1\\
\hline
$<{\bf X}|\psi _4>$&[4,0,0,0]&(4,0,0,0)&$(G_2)^4$&8&$2^4$&105\\
$ $&[3,1,0,0]&(2,1,0,0)&$(G_2)^2G_3$&7&$2^23^1$&105\\
$ $&[2,1,1,0]&(1,0,1,0)&$G_2G_4$&6&$2^14^1$&15\\
$ $&[2,2,0,0]&(0,2,0,0)&$(G_3)^2$&6&$3^2$&10\\
$ $&[1,1,1,1]&(0,0,0,1)&$G_5$&5&$5^1$&1\\
\hline
\end{tabular}
\end{center}
\end{table}
}
\vspace*{0.5cm}

\small{\subsection{Mandelstam identities of (I) and (II)}}
\normalsize{
Let us consider the first and second constraints of the Mandelstam identities. From Eq.(5) and Eq.(6), we obtain the following relations for the basic operator $G_r$:
\begin{enumerate}
\item[(I')]
\begin{eqnarray}
<\mu_1\cdots \mu_r|G_{r}|0>=<(\mu_1\cdots \mu_r)_c|G_{r}|0>.
\end{eqnarray}
\item[(II')]
\begin{eqnarray}
<\mu_1\cdots \mu_r|G_{r}|0>=(-1)^r<\mu_r\cdots \mu_1|G_{r}|0>.
\end{eqnarray}
\end{enumerate}
From Eq.(22), the function $G_{\mu_r\cdots \mu_1}$ is cyclic symmetric. Thus, it satisfies Eq.(31). From Eq.(32), we get 
\begin{eqnarray}
G_{\mu_1\cdots \mu_r}=(-1)^rG_{\mu_r\cdots \mu_1}.
\end{eqnarray}
For $r=2$, $G_{\mu_1\mu_2}$ is symmetric and for $r=3$, $G_{\mu_1\mu_2\mu_3}$     is antisymmetric.
}

\small{\section{Quantum states based on Chern-Simons theory}}
\small{\subsection{Chern-Simons type propagators }}
\normalsize{
 In quantum gravity, the extended loop wave function $\psi{({\bf X})}$ must satisfy the Mandelstam identities, the diffeomorphism constraint and the Hamiltonian constraint. Let us consider the solutions to satisfy these constraints. 
In the extended loop calculus \cite{J. Griego,J. GriegoB467} based on the Chern-Simons theory, the propagator are written by the products of the two-point propagator$(g_{\mu_1\mu_2})$ and the three-point propagators$( h_{\mu_1\mu_2\mu_3})$, given in Appendix A. These propagators satisfy the properties of $G_r$ discussed in the previous section, Eq.(31) and Eq.(32). Thus, we assume that the operator $\tilde{G}$ is built up in terms of the following basic propagator operators for $G_2$ and $G_3$:

(a) the propagator operator with the rank 2($ G_2=g$)
\begin{eqnarray}
<\mu_d|g|0>=\delta_{\mu_d}^{(\mu_1\mu_2)}g_{\mu_1\mu_2},
\end{eqnarray}

(b) the propagator operator with the rank 3($G_3=h$)
\begin{eqnarray}
<\mu_d|h|0>=\delta_{\mu_d}^{(\mu_1\mu_2\mu_3)}h_{\mu_1\mu_2\mu_3}.
\end{eqnarray}

Furthermore, from the analysis for the third coefficient of the Jones polynomial \cite{C. Di Bartolo}, we assume the following basic propagator with the rank 4($G_4=[hg^{-1}h]$)
:
\begin{eqnarray}
<\mu_d|hg^{-1}h|0>=\delta_{\mu_d}^{(\mu_1\mu_2\mu_3\mu_4)}G_{\mu_1\mu_2\mu_3\mu_4}
\end{eqnarray}
with
\begin{eqnarray}
G_{\mu_1\mu_2\mu_3\mu_4}=h_{\alpha (\mu_1\mu_2}g^{\alpha \beta }h_{\mu_3\mu_4)_c\beta }\nonumber
\end{eqnarray}
where $g^{-1}$ denotes the inverse operator of $g$. The function $g^{\alpha \beta}$ is the inverse function $g_{\alpha \beta}$ and indicates an inner propagator. The index $c$ denotes the sum under cyclic permutation with respect to $\mu$. It is possible to generalize the form of Eq.(36) to $G_5$. The propagator operator with the rank 5($G_5= [h(g^{-1}h)^2]$) is given as
\begin{eqnarray}
<\mu_d|hg^{-1}hg^{-1}h|0>=\delta_{\mu_d}^{(\mu_1\cdots \mu_5)}G_{\mu_1\cdots \mu_5}
\end{eqnarray}
with
\begin{eqnarray}
G_{\mu_1\cdots \mu_5}=h_{\alpha (\mu_1\mu_2}g^{\alpha \beta }h_{\beta \mu_3\gamma }g^{\gamma \delta }h_{\mu_4\mu_5)_c\delta}.\nonumber
\end{eqnarray}
These basic propagators satisfy the cyclic symmetry of Eq.(5) and the inverse symmetry of Eq.(6). The propagators with the rank $n(n \ge 6)$ are similarly defined.
}
\vspace*{0.5cm}

\small{\subsection{Vacuum quantum states}}
\normalsize{ 
Let us investigate the extended loop wave function with the family number $m=1,2 $ and $3$ in the case of the vacuum Hamiltonian.

First we proceed with the case of $m=1$. The partition of 1 is $\{[1]\}$. The extended loop wave function with $m=1$ is given as
\begin{eqnarray}
<{\bf X}|\psi_1>=X^{\mu_1 \mu_2}D_{\mu_1 \mu_2}^{(\mu_1\mu_2)}<(\mu_1\mu_2)|g|0>=\alpha _{0}^{1}{\varphi}_G
\end{eqnarray}
with
\begin{eqnarray}
{\varphi}_G=X^{\mu_1 \mu_2}g_{\mu_1 \mu_2}\nonumber
\end{eqnarray}
where $\alpha _{0}^{1}$ denotes a coefficient. Thus, the propagator $\psi_{\mu_1 \mu_2}^1$ is given as $g_{\mu_1 \mu_2}$. 
This propagator satisfies the Mandelstam identities  (I), (II) and (III) \cite{D. Shao}. 
Also, $<{\bf X}|\psi_1>$ satisfies the diffeomorphism constraint but do not the Hamiltonian constraint as discussed in \cite{D. Shao}.

Secondly, we treat the case of $m=2$. The partition of 2 is $\{[2,0], [1,1]\}$ and hence the wave function with $m=2$ is given as
\begin{eqnarray}
<{\bf X}|\psi_2>=X^{\mu_1\cdots  \mu_4}\psi_{\mu_1\cdots  \mu_4}^{2}+X^{\mu_1\cdots  \mu_3}\psi_{\mu_1\cdots  \mu_3}^{2}
\end{eqnarray}
with
\begin{eqnarray}
\psi_{\mu_1\cdots  \mu_4}^{2}&=&\alpha _1^4g_{\mu_1 \mu_2}g_{\mu_3 \mu_4}+\alpha _2^4g_{\mu_1 \mu_3}g_{\mu_2 \mu_4}+\alpha _3^4g_{\mu_1 \mu_4}g_{\mu_2 \mu_3}\nonumber\\
\psi_{\mu_1\cdots  \mu_3}^{2}&=&\alpha _0^3\nonumber
\end{eqnarray}
where $\alpha_i^j$ denotes a expansion coefficient. 

Eq.(39) satisfies the Mandelstam identity (II). While, from the constraint of the Mandelstam identity (I), we obtain $\alpha_1^4=\alpha_3^4$. Thus, the number of independent parameters is three. We select $(\alpha_1^4,\alpha_2^4,\alpha_0^3)$. Furthermore, we require the diffeomorphism constraint. Then, we obtain the following equation:
\begin{eqnarray}
\alpha_1^4-\alpha_2^4+\alpha_0^3=0. 
\end{eqnarray}
As the independent solutions, we select the following two solutions:

a) $\alpha _0^3=0$ and $\alpha _1^4=\alpha _2^4=1$
\begin{eqnarray}
\psi_2^1({\bf X})=X^{\mu_1\cdots  \mu_4}(g_{\mu_1 \mu_2}g_{\mu_3 \mu_4}+g_{\mu_1 \mu_3}g_{\mu_2 \mu_4}+g_{\mu_1 \mu_4}g_{\mu_2 \mu_3})\equiv \frac{1}{2!}(\varphi_G)^2.
\end{eqnarray}

b) $\alpha _1^4=0$ and $\alpha _2^4=\alpha _0^3=1$
\begin{eqnarray}
\psi_2^2({\bf X})=X^{\mu_1\cdots  \mu_4}(g_{\mu_1 \mu_3}g_{\mu_2 \mu_4}+X^{\mu_1\mu_2 \mu_3}h_{\mu_1\mu_2 \mu_3})\equiv J_2.
\end{eqnarray}
where $J_2$ denotes the second coefficient of the Jonse polynomial. We obtain the two-parameter family of solutions as
\begin{eqnarray}
\psi_2(a,b)=a\psi_2^1({\bf X})+b\psi_2^2({\bf X}).
\end{eqnarray}
where $a$ and $b$ are parameters. For examples, we have
\begin{eqnarray}
& &\psi_2(1,-1)=\frac{1}{2!}(\varphi_G)^2+J_2,\ \ \psi_2(1,1)=\frac{1}{2!}(\varphi_G)^2+J_2,\nonumber\\
& &\psi_2(2,-1)=\frac{1}{2!}(\varphi_G)^2+K_2.
\end{eqnarray}
where $K_2$ denotes the second coefficient of the Kauffman bracket and $K_2=\psi_2(1,-1)$. It is possible to show that these quantum states satisfy the Mandelstam identities (I), (II), and (III) \cite{D. Shao}. The set of $\{$$\psi_2(0,1), \psi_2(1,-1)$$\}$ corresponds to the extended knot family $\{\psi_i\}_3^4$  given by Griego in \cite{J. Griego}. 
While the Hamiltonian constraint gives us the relation of $\alpha _1^4= 0$ and $\alpha_2^4 =\alpha_0^3$. This solution corresponds to the state of $J_2$. The quantum state $J_2$ satisfies the all constraints.

Finally, we treat the case of $m=3$. Then, the partition of $3$ is given as follows:$\{$ $[3,0,0], [2,1,0]$,$ [1,1,1]$$\}$. Thus, the wave function with $m=3$ is given as 
\begin{eqnarray}
<{\bf X}|\psi_3>=X^{\mu_1\cdots  \mu_6}\psi_{\mu_1\cdots  \mu_6}^{3}+X^{\mu_1\cdots  \mu_5}\psi_{\mu_1\cdots  \mu_5}^{3}+X^{\mu_1\cdots  \mu_4}\psi_{\mu_1\cdots  \mu_4}^{3}
\end{eqnarray}
with
\begin{eqnarray}
\psi_{\mu_1\cdots  \mu_6}^{3}&=&\sum_{i=1}^{15}\alpha _i^6\phi _i^6,\ \ \ \psi_{\mu_1\cdots  \mu_5}^{3}=\sum_{i=1}^{10}\alpha _i^5\phi _i^5,\nonumber\\
\psi_{\mu_1\cdots  \mu_4}^{3}&=&\alpha _0^4\phi _0^4
\end{eqnarray}
where $\phi_i^6$, $\phi_i^5$ and $\phi_i^4$ are given in Appendix B. The number of parameters is $26$.
 
The constraint of the Mandelstam identity (I) for Eq.(45) yields the following relations for $\alpha _i^j$:
\begin{eqnarray}
& &\alpha _1^6=\alpha _2^6,\ \  \alpha_3^6 =\alpha_4^6 =\alpha_5^6, \nonumber\\
& &\alpha _6^6=\alpha _7^6=\alpha_8^6=\alpha_9^6 =\alpha_{10}^6=\alpha _{11}^6,\ \  \alpha_{12}^6 =\alpha_{13}^6=\alpha _{14}^6,\nonumber\\
& & \alpha _1^5=\alpha _2^5,\ \  \alpha_3^5 =\alpha_4^5=\alpha_5^5,\ \  \alpha _6^5=\alpha _7^5=\alpha_8^5=\alpha_9^5 =\alpha_{10}^5.
\end{eqnarray}
While, the Mandelstam identity (II) requires the following relations for $\alpha _i^j$: 
\begin{eqnarray}
& &\alpha _3^6=\alpha _4^6,\ \  \alpha_6^6 =\alpha_{10}^6,\ \  \alpha_7^6=\alpha_9^6,\ \  \alpha_{13}^6=\alpha _{14}^6,\nonumber\\
& &\alpha _1^5=\alpha _5^5,\ \  \alpha_3^5 =\alpha_4^5,\ \  \alpha _6^5=\alpha _{10}^5,\ \  \alpha_7^5 =\alpha_{9}^5.
\end{eqnarray}
These relations are contained in Eq.(47). Therefore they do not give the further constraint. The number of parameters is 8. We select $\alpha _1^6$, $\alpha _3^6$, $\alpha_6^6$, $\alpha_{12}^6$, $\alpha _{15}^6$, $\alpha _{1}^5$, $\alpha_6^5$ and  $\alpha_{0}^4$.

In terms of adding the diffeomorphism constraint to the condition of the Mandelstam identity (I), we obtain the following five equations: 
\begin{eqnarray}
& &\alpha_1^6-\alpha_3^6=0,\ \ \ \ \alpha_{12}^6-\alpha_{15}^6+\alpha_6^5=0,\nonumber\\
& &\alpha_1^6-\alpha_6^6+\alpha_1^5=0,\ \ \ \ \alpha_{6}^6-\alpha_{12}^6+\alpha_6^5=0,\nonumber\\
& &\alpha_1^5-\alpha_6^5+\alpha_0^4=0. 
\end{eqnarray}
Hence, the number of the independent solutions is three. We select the solutions given in Table \ref{tt:2}. The functions $J_3$ and $K_3$ correspond to the third coefficients of the Jones polynomial and the third coefficients of the Kauffman bracket, respectively.
\begin{table}[b]
\caption{The wave functions $\psi_3^1$, $\psi_3^2$ and $\psi_3^3$}
\label{tt:2}
\begin{center}
\begin{tabular}{|c| c c c c c c c c|c|}
\hline
$\psi_3^i$&$\alpha_1^6$&$\alpha_3^6$&$\alpha_6^6$&$\alpha_{12}^6$&$\alpha_{15}^6$&$\alpha_1^5$&$\alpha_6^5$&$\alpha_0^4$&Knot invariant\\
\hline
$\psi_3^1$&1&1&1&1&1&0&0&0&{$(\varphi_G)^3/3!$}\\
$\psi_3^2$&0&0&0&1&2&0&1&1&{$J_3$}\\
$\psi_3^3$&1&1&0&0&0&-1&0&1&{$K_3$}\\
\hline
\end{tabular}
\end{center}
\end{table}
Thus, we obtain the three-parameter family of solutions as
\begin{eqnarray}
\psi_3(a,b,c)=a\psi_3^1({\bf X})+b\psi_3^2({\bf X}+c\psi_3^3({\bf X}).
\end{eqnarray}
where $a$, $b$ and $c$ are parameters. 

For examples, we have
\begin{eqnarray}
& &\psi_3(1,-1,0)=\frac{1}{3!}(\varphi_G)^3-J_3,\ \ \psi_3(1,0,-1)=\frac{1}{3!}(\varphi_G)^3-K_3,\nonumber\\
& &\psi_3(1,1,0)=\frac{1}{3!}(\varphi_G)^3+J_3,\ \ \psi_3(1,0,1)=\frac{1}{3!}(\varphi_G)^3+K_3,\nonumber\\
& &\psi_3(1,1,-1)=\frac{1}{3!}(\varphi_G)^3+J_3-K_3\equiv {\varphi_G}J_2.
\end{eqnarray}
The set of $\{$$\psi_3(0,0,1)$, $\psi_3(1,0,-1)$, $\psi_3(0,1,0)$, $\psi_3(1,-1,0)$$                                         \}$ corresponds to the extended knot family $\{\psi\}_4^6$  in \cite{J. Griego}. While, the Hamiltonian constraint gives the relations $\alpha _1^6=\alpha _3^6=\alpha _6^6=\alpha _{12}^6=\alpha _{15}^6=\alpha _1^5=\alpha _6^5=\alpha _6^4=0$. It implies that Eq.(50) does not satisfy the Hamiltonian constraint. Also, it is possible to show that theses quantum states do not satisfy the Mandelstam identity (III) \cite{L. Shao}.
}
\vspace*{0.5cm}

\small{\subsection{Quantum state with $\Lambda$}}
\normalsize{    
We may apply the operator formalism to the case of the Hamiltonian with $\Lambda$. We consider only the operator $G_2=g$ and use Eq.(18). Then, we have
\begin{eqnarray}
\psi_0({\bf X})&=&\sum_{m=0}^{\infty }{\lambda}^m X^{\mu_1\cdots \mu_{2m}}\sum_{\tilde{P}_{2m}{[(\mu_1\mu_2)\cdots (\mu_{2m-1}\mu_{2m})]}}\nonumber\\
&\times  &D_{\mu_1\cdots \mu_{2m}}^{(\mu_1\mu_2)\cdots (\mu_{2m-1}\mu_{2m})}<(\mu_1\mu_2)\cdots (\mu_{2m-1}\mu_{2m})|g^m|0>.
\end{eqnarray}
If $D_{\mu_1\cdots \mu_{2m}}^{(\mu_1\mu_2)\cdots (\mu_{2m-1}\mu_{2m})}=1$, we obtain the following relation:
\begin{eqnarray}
\frac{1}{m!}(\varphi_G)^m&=&\sum_{\tilde{P}_{2m}{[(\mu_1\mu_2)\cdots (\mu_{2m-1}\mu_{2m})]}}X^{\mu_1\cdots \mu_{2m}}\nonumber\\
&\times& <(\mu_1\mu_2)\cdots (\mu_{2m-1}\mu_{2m})|g^m|0>.
\end{eqnarray}
Hence, we obtain the following wave function:
\begin{eqnarray}
\psi_0({\bf X})=e^{\frac{\Lambda}{6}a_1}
\end{eqnarray}
where $\lambda= -\Lambda/4$ and $a_1=-3\varphi_G/2$. The wave function of Eq.(54) indicates the phase factor of the Kauffman bracket($\psi_{AK}$). Eq.(54) satisfies the diffeomorphism and Hamiltonian constraints with $\Lambda$ as discussed in \cite{D. Shao03}. From Eq.(20), this wave function is compactly rewritten as
\begin{eqnarray}
\psi_0({\bf X})=<{\bf X}|(1+\frac{\Lambda}{4}g)^{-1}|0>
\end{eqnarray}
}

\small{\section{Conclusion and discussion}}
\normalsize{
We have investigated the structure of the extended loop wave functions by using the operator formalism. We have shown that the extended loop wave functions have the following structure:
\begin{enumerate} 
\item[1)] The extended loop wave functions are characterized by the family number $m$.
\item[2)] The extended loop wave function with m, is decomposed into the classes in terms of the partition of $m$; [$\tau $]=[$\tau _1,\cdots ,\tau_m$]. 
\item[3)] The extended loop wave functions are structured according a sequential arrangement of basic propagators.
\item[4)] The extended loop wave function exhibits a skelton structure.
\end{enumerate}

 In quantum gravity, we require the constraints for the extended loop wave functions constructed by our approach. We have found that these constraints exhibit the following hierarchy structure:
 \begin{enumerate}  
\item[(1)]	The Mandelstam identity (II),
\item[(2)]	The Mandelstam identity (I),
\item[(3)]	Diffeomorphism constraint,
\item[(4)]	The Mandelstam identity (III),
\item[(5)]	Hamiltonian constraint.
\end{enumerate}
The strongest condition is the Hamiltonian constraint. It is well-known that in the case of the vacuum Hamiltonian, the solution to satisfy all constraints is only $J_2$ until now.

 Also, our analysis suggests that the number of the independent solution in the extended loop wave function with $m$ equals to the number of the class of the partition of $m$. Thus it is necessary to study the extended loop wave function with $m=4$. This point is a chalenging problem and will be investigated elswhere.  

\vspace*{2cm}

{\bf Appendix A}\ \    The propagators in the Chern-Simons theory

\vspace*{0.5cm}

The two-point propagator $g_{\mu_1\mu_2}$ and three-point propagators $h_{\mu_1\mu_2\mu_3}$ are defined as follows:
\begin{eqnarray}
g_{\mu_1\mu_2}=g_{axby}=-\frac{\epsilon _{abc}}{4\pi }\frac{(x-y)^c}{|x-y|^3},\nonumber\\
h_{\mu_1\mu_2\mu_3}=\epsilon ^{\alpha_1 \alpha_2 \alpha_3 }g_{\mu_1\alpha_1}g_{\mu_2\alpha_2}g_{\mu_3\alpha_3}
\end{eqnarray}
with
\begin{eqnarray}
\epsilon ^{\alpha_1 \alpha_2 \alpha_3 }=\epsilon ^{b_1y_1b_2y_2cx }=\epsilon ^{b_1b_2c }\delta(x-y_1)\delta(x-y_2).
\end{eqnarray}

When the skeleton structure of the extended loop wave function is considered, the following inner propagator $g^{\alpha \beta }$ plays the important role:
\begin{eqnarray}
g^{\alpha \beta}=g^{axby}=\epsilon ^{b_1b_2c }\partial _c^x\delta(x-y).\nonumber
\end{eqnarray}
where $g^{\alpha \beta}$  is the inverse of the two-point propagator of the Chern-Simons theory and is defined in the space of the transverse function. It satisfies the following relation:
\begin{eqnarray}
g^{\alpha \beta}g_{\beta \gamma }=g^{ax\beta }g_{\beta cz}=(\delta_c^a-\partial ^a\partial _c\nabla ^{-2})\delta(x-y).\nonumber
\end{eqnarray}

\vspace*{0.5cm}

{\bf Appendix B}\ \   The functions $\phi _i^6$, $\phi_i^5$ and $\phi_0^4$

\vspace*{0.5cm}

1) $\phi _i^6(i=1\sim 15)$
\begin{eqnarray}
& &\phi_1^6=g_{\mu_1 \mu_2}g_{\mu_3 \mu_4 }g_{\mu_5 \mu_6},\ \ \phi_2^6=g_{\mu_1 \mu_6}g_{\mu_2 \mu_3 }g_{\mu_4 \mu_5},\ \ \phi_3^6=g_{\mu_1 \mu_2}g_{\mu_3 \mu_6 }g_{\mu_4 \mu_5},\nonumber\\
& &\phi_4^6=g_{\mu_1 \mu_4}g_{\mu_2 \mu_3 }g_{\mu_5 \mu_6},\ \ \phi_5^6=g_{\mu_1 \mu_6}g_{\mu_2 \mu_5 }g_{\mu_3 \mu_4},\ \ \phi_6^6=g_{\mu_1 \mu_2}g_{\mu_3 \mu_5 }g_{\mu_4 \mu_6},\nonumber\\
& &\phi_7^6=g_{\mu_1 \mu_5}g_{\mu_2 \mu_3 }g_{\mu_4 \mu_6},\ \ \phi_8^6=g_{\mu_1 \mu_5}g_{\mu_3 \mu_4 }g_{\mu_2 \mu_6},\ \ \phi_9^6=g_{\mu_1 \mu_3}g_{\mu_2 \mu_6 }g_{\mu_4 \mu_5},\nonumber\\
& &\phi_{10}^6=g_{\mu_1 \mu_3}g_{\mu_2 \mu_4 }g_{\mu_5 \mu_6},\ \ \phi_{11}^6=g_{\mu_1 \mu_6}g_{\mu_2 \mu_4 }g_{\mu_3 \mu_5},\ \ \phi_{12}^6=g_{\mu_1 \mu_3}g_{\mu_2 \mu_5 }g_{\mu_4 \mu_6},\nonumber\\
& &\phi_{13}^6=g_{\mu_1 \mu_5}g_{\mu_2 \mu_4 }g_{\mu_3 \mu_6},\ \ \phi_{14}^6=g_{\mu_1 \mu_5}g_{\mu_2 \mu_6 }g_{\mu_3 \mu_5},\nonumber\\
& &\phi_{15}^6=g_{\mu_1 \mu_4}g_{\mu_2 \mu_5 }g_{\mu_3 \mu_6}.\nonumber
\end{eqnarray}

2) $\phi_i^5(i=1\sim 10)$
\begin{eqnarray}
& &\phi_1^5=g_{\mu_1 \mu_2}h_{\mu_3 \mu_4 \mu_5},\ \ \phi_2^5=g_{\mu_1 \mu_5}h_{\mu_2 \mu_3 \mu_4},\ \ \phi_3^5=g_{\mu_2 \mu_3}h_{\mu_1 \mu_4 \mu_5},\nonumber\\
& &\phi_4^5=g_{\mu_3 \mu_4}h_{\mu_1 \mu_2 \mu_4},\ \ \phi_5^5=g_{\mu_4 \mu_5}h_{\mu_1 \mu_2 \mu_3},\ \ \phi_6^5=g_{\mu_1 \mu_3}h_{\mu_2 \mu_4 \mu_5},\nonumber\\
& &\phi_7^5=g_{\mu_1 \mu_4}h_{\mu_2 \mu_3 \mu_5},\ \ \phi_8^5=g_{\mu_2 \mu_4}h_{\mu_1 \mu_3 \mu_5},\ \ \phi_9^5=g_{\mu_2 \mu_5}h_{\mu_1 \mu_3 \mu_4},\nonumber\\
& &\phi_{10}^5=g_{\mu_3 \mu_5}h_{\mu_1 \mu_2 \mu_4}.\nonumber
\end{eqnarray}

3) $\phi_0^4$
\begin{eqnarray}
\phi_0^4=h_{\alpha(\mu_1\mu_2}{g^{\alpha \beta} }_{\mu_3\mu_4)_c\beta}.\nonumber
\end{eqnarray}

}


\begin{thebibliography}{99}
\bibitem{Ashtk86} A. Ashtekar, $Phys.\ Rev.\ Lett.\ {\bf 57}, $ 2244 (1986), $Phys.\ Rev.\ {\bf D36}$, 1587 (1987).
\bibitem{Ashtk90} A. Ashtekar, Lectures on non-perturbative quantum gravity, editors R. Kulkarui, J. Samel, University of Poona Press, Poona (1990).

H. Kodama, $Phys.\ Rev.\ {\bf D42 }$, 2548 (1990).

\bibitem{E. Guadagnini}E. Guadagnini, M. Martellini and M. Mintchev, $Nucl.\ Phys.\ {\bf B330}$, 575 (1990).

B. Br{\"u}gmann, R. Gambini and J. Pullin, $Nucl.\ Phys.\ {\bf B385}$, 587 (1992).
\bibitem{J. Griego}J. Griego, $Nucl.\ Phys.\ {\bf B473}$, 291 (1996).
\bibitem{D. Shao}D. Shao, H. Noda, L. Shao and C. G. Shao, $Int.\ J.\ Mod.\ Phys.\ {\bf 17}$, 615 (2002).
\bibitem{S. Mandelstam}S. Mandelstam, $Phys.\ Rev.\ {\bf D19}$, 2391 (1979)
\bibitem{C. Di Bartolo94}C. Di Bartolo, R. Gambini, J. Griego and J. Pullin, $Phys.\ Rev.\ Lett.\ {\bf 72}, $ 3638 (1994).
\bibitem{C. Di Bartolo95}C. Di Bartolo, R. Gambini and J. Griego, $Phys.\ Rev.\ {\bf D51}$,  502 (1995).
\bibitem{J. GriegoB467}J. Griego, $Nucl.\ Phys.\ {\bf B467}$, 332 (1996).
\bibitem{D. Shao03}D. Shao, H. Noda, L. Shao, M. Saito and C. G. Shao, Cosmological term and extended loop representation of quantum gravity(2003).
\bibitem{C. Di Bartolo}C. Di Bartolo and J. Griego, $ Phys.\ Lett.\ {\bf B317}$, 540 (1993).
\bibitem{L. Shao}L. Shao, H. Noda, D. Shao and C. G. Shao, Calculation of Hamiltonian constraint and Mandelstam identities over Extended knot family $\{\psi_i \}_4^6$ (2002).
\end{thebibliography}
\end{document}